\documentstyle[12pt,epsf]{article}
\begin{document}

\begin{titlepage}

\begin{flushright}
PSI-PR-97-21
\\ \today  \\[20mm]
\end{flushright}

\begin{center}

{\Large\bf A Precision Measurement of \\[2mm]
           Nuclear Muon Capture on $^3$He}

\end{center}

\begin{center}

P. Ackerbauer$^{1}$,
D.V. Balin$^{2}$,
V.M. Baturin$^{2}$,
G.A. Beer$^{8}$,
\\W.H. Breunlich$^{1}$,
T. Case$^{6}$,
K. Crowe$^{6}$,
H. Daniel$^{5}$,
J. Deutsch$^{3}$,
J. Govaerts$^{3}$,
\\Yu.S. Grigoriev$^{2}$,
F.J. Hartmann$^{5}$,
P. Kammel$^{6}$,
R. King$^{1}$,
B. Lauss$^{1}$,
E.M. Maev$^{2}$,
\\V.E. Markushin$^{4,7}$,
J. Marton$^{1}$,
M. M\"uhlbauer$^{5}$,
C. Petitjean$^{4}$,
Th. Petitjean$^{4}$,
\\G.E. Petrov$^{2}$,
R. Prieels$^{3}$,
W. Prymas$^{1}$\renewcommand{\thefootnote}{\fnsymbol{footnote}}\footnote[1]
{This paper is part of the thesis works of W. Prymas and N.I. Voropaev},
W. Schott$^{5}$,
G.G. Semenchuk$^{2}$,
\\Yu.V. Smirenin$^{2}$,
A.A. Vorobyov$^{2}$,
N.I. Voropaev$^{2*}$,
P. Wojciechowski$^{5}$ 
\vspace*{5mm}\\
{\small\sl
(1) Austrian Academy of Sciences\\
Institute for Medium Energy Physics,
 A-1090 Wien, Austria\\
(2) Petersburg Nuclear Physics Institute (PNPI), Gatchina 188350, Russia\\
(3) Catholic University of Louvain, B-1348 Louvain-la-Neuve, Belgium\\
(4) Paul Scherrer Institute, CH-5232 Villigen, Switzerland\\
(5) Technical University of Munich, D-85747 Garching, Germany\\
(6) University of California at Berkeley and\\
Lawrence Berkeley Laboratory, 94729 California, USA\\
(7) Kurchatov Institute, 123182 Moscow, Russia\\
(8) University of Victoria, Victoria B.C. V8W 2Y2, Canada\\
}
\end{center}   

\begin{abstract}

The muon capture rate in the reaction
$\mu^-~+~^{3}{\rm He}~\rightarrow \nu_\mu~+~^{3}{\rm H}$
has been measured at PSI using a modular high pressure
ionization chamber. The rate corresponding to statistical hyperfine population
of the $\mu^3$He atom is ($1496.0 \pm 4.0)\,\rm{s}^{-1}$.
This result  confirms the PCAC prediction for the pseudoscalar form factors
of the $^3$He~-$^3$H system and the nucleon. 
\\[4mm]

\end{abstract}



\end{titlepage}

\section{Introduction}

We report on a measurement of the muon-$^3$He capture rate to the triton
channel, $\mu^-~+~^{3}{\rm He}~\rightarrow \nu_\mu~+~^{3}{\rm H}\ (1.9\
{\rm MeV})$, performed to a level of precision unprecedented in nuclear
muon capture experiments.

The main information gained from this measurement is the determination of the
induced nuclear pseudoscalar form factor $F_{\rm P}(q^2=-0.954 m_{\mu}^2)$
in the $A=3$ system treated in the frame of the Elementary Particle Model (EPM)
pioneered by Kim and Primakoff\cite{Prim65} and elaborated recently by
Congleton and Fearing\cite{Con92}.
The value obtained for $F_{\rm P}$ provides a
stringent test of the Partially Conserved Axial Current (PCAC)
approximation\cite{Adl66}, based on the Goldstone mode realization of the
approximate chiral symmetries of QCD.

In addition, comparison of the measured capture rate with that calculated
according to the presently available\cite{Con96} microscopic model of muon
capture by $^3$He allows a test of the validity of this approach which
takes the
Meson Exchange Currents (MEC) explicitly into account. Using this model,
one can evaluate the induced pseudoscalar form factor of the nucleon,
$g_{\rm P}(q^2=-0.88 m_{\mu}^2)$. Furthermore, our experimental result also
allows for further tests of the Standard Model in its electroweak
sector\cite{Gov97}.

In spite of the uncertainties in the various form factors entering the
calculation, PCAC and EPM predict the capture rate to a precision
of 1.4\%\cite{Con92}. Until now the experimental situation  did not equal such
a precision, since only three measurements\cite{Fal63,Aue65,Cla65}, 
dating back to over thirty years ago were available,
with precisions ranging from 3\% to 10\%.
The situation thus called for a new precision experiment which is described
in this work. Preliminary results were published earlier\cite{Bal95,Vor96}.

\section{Experiment}

One of the main advantages in measuring nuclear muon capture on $^{3}$He, 
compared to hydrogen, is the production 
of a charged particle in the final state, which can be detected with 
high efficiency and good background suppression.
For muon capture on $^3$He,
the main reaction channel is
\begin{equation}
\mu^- + ^{3}\!{\rm He} \longrightarrow ^{3}\!{\rm H} + \nu_\mu\ \
(70\%), \\
\label{eq:reac1}
\end{equation}
accompanied by triton breakup into $d + n$ (20\%) and $p + 2n$
(10\%)\cite{Phi75}.

\begin{figure}[hbp]
\newsavebox{\muHe}
\savebox{\muHe}(0,0)[cc]{
\put(0,50){\circle{40}}
\put(-15,50){$\mu^3$He}
\put(-15,40){$\ _{F=0}$}
\put(20,50){\line(1,0){50}}
\put(70,50){\line(3,1){60}}
\put(70,50){\line(3,-1){60}}
\put(20,50){\vector(1,0){25}}
\put(70,50){\vector(3,1){30}}
\put(70,50){\vector(3,-1){30}}
\put(45,55){$\lambda^{0}_{c}$}
\put(100,70){$\lambda^{0}_{\rm H}$}
\put(100,25){$\lambda^{0}_{n}$}
\put(135,70){$\nu_{\mu}+^3$H}
\put(135,33){$\nu_{\mu}+d+n$}
\put(135,17){$\nu_{\mu}+p+2n$}
\put(15,65){\line(1,1){25}}
\put(15,65){\vector(1,1){11}}
\put(30,95){$e^{-}+\bar{\nu}_e+\nu_{\mu}+^3$He}
\put(15,80){$\lambda_0$}
\put(0,-50){\circle{40}}
\put(-15,-50){$\mu^3$He}
\put(-15,-60){$\ _{F=1}$}
\put(20,-50){\line(1,0){50}}
\put(70,-50){\line(3,1){60}}
\put(70,-50){\line(3,-1){60}}
\put(20,-50){\vector(1,0){25}}
\put(70,-50){\vector(3,1){30}}
\put(70,-50){\vector(3,-1){30}}
\put(45,-45){$\lambda^{1}_{c}$}
\put(100,-30){$\lambda^{1}_{\rm H}$}
\put(100,-75){$\lambda^{1}_{n}$}
\put(135,-30){$\nu_{\mu}+^3$H}
\put(135,-67){$\nu_{\mu}+d+n$}
\put(135,-83){$\nu_{\mu}+p+2n$}
\put(15,-65){\line(1,-1){25}}
\put(15,-65){\vector(1,-1){10}}
\put(30,-100){$e^{-}+\bar{\nu}_e+\nu_{\mu}+^3$He}
\put(15,-88){$\lambda_0$}
\put(0,30){\line(0,-1){60}}
\put(0,30){\vector(0,-1){30}}
\put(10,0){$\lambda^{0\to 1}$}
\put(-100,0){\circle{20}}
\put(-103,-2){$\mu$}
\put(-90,7){\line(2,1){70}}
\put(-90,-7){\line(2,-1){70}}
\put(-90,7){\vector(2,1){34}}
\put(-90,-7){\vector(2,-1){34}}
\put(-70,30){$1/4$}
\put(-70,-37){$3/4$}
}
\mbox{
     \begin{picture}(240,200)(-120,-100)
        \put(0,0){\usebox{\muHe}} 
     \end{picture}
     }
\vspace*{10mm} 
\caption{\label{FigScheme} 
Kinetics scheme of the $\mu^3$He system.
The main disappearance of the muon is by decay with the rate
$\lambda_0$ (99.7\%).
The total capture rate from the hyperfine structure state $F=0,1$
is $\lambda^F_c=\lambda^F_{\rm H}+\lambda^F_{n}$, where  
$\lambda^F_{\rm H}$ is the capture rate to the
$\nu_{\mu}+^3\!\mbox{\rm H}$ channel and
$\lambda^F_{n}$ is the capture rate to the
$\nu_{\mu}+d+n$ and $\nu_{\mu}+p+2n$ channels. 
}
\end{figure}
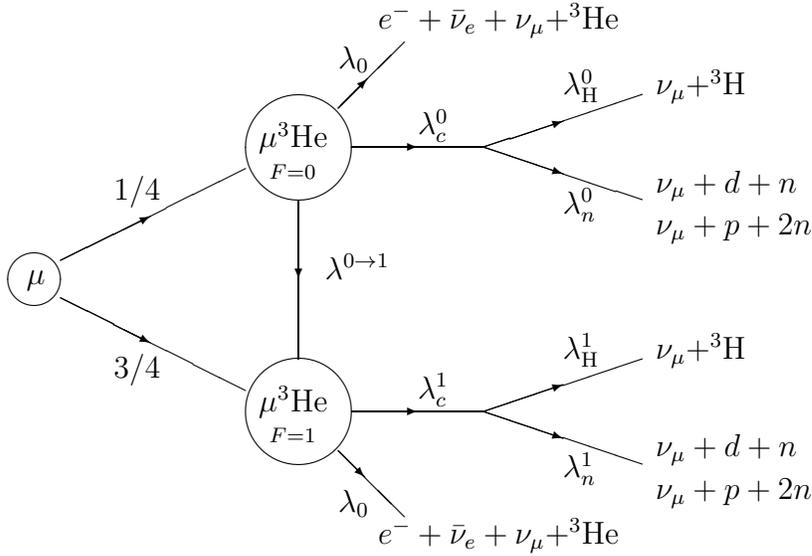

These reactions occur from the two hyperfine states of 
the $\mu^{3}$He muonic atom, of total spin $F=0$ and $F=1$,
see the kinetics scheme in Fig.\ref{FigScheme}.
Since in the present experiment the $^3$He target is not polarized and the
spin flip is small (cf. results below), 
the hyperfine states are statistically populated, and
it is the statistical capture rate to the triton channel
\begin{equation} 
\lambda_{\rm stat} = \frac{1}{4}\lambda^0_{\rm H}
                    +\frac{3}{4}\lambda^1_{\rm H}
\label{lambdastat}
\end{equation}
which is measured\cite{Con92,Con93}.

\begin{figure}[hbp]
\begin{center} \mbox{\epsfysize=80mm\epsffile{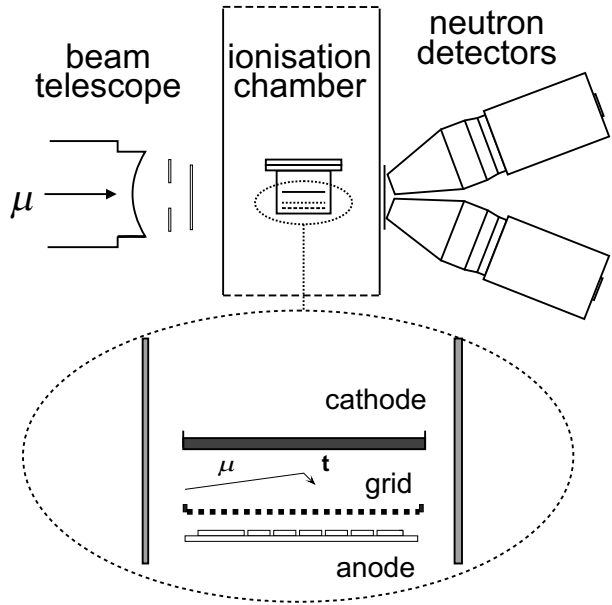}} \end{center}
\caption{\label{FigSetup} 
Experimental setup and layout of the ionization chamber.} 
\end{figure}

\begin{figure}[htb]
\begin{center} \mbox{\epsfysize=80mm\epsffile{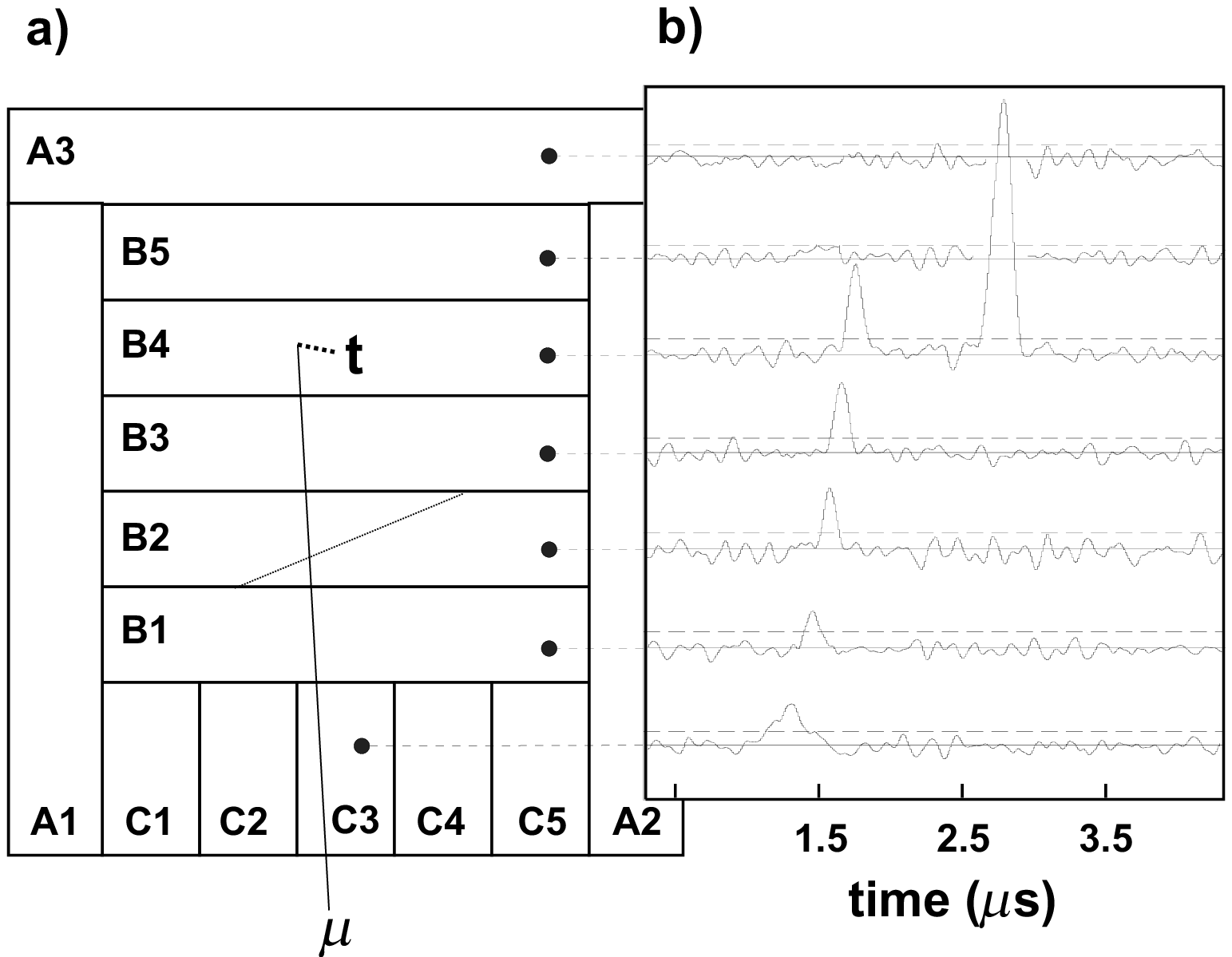}} \end{center}
\caption{\label{FigAnode} 
(a) Anode layout of the ionization chamber. 
(b) A typical sequence of anode signals registered by the flash ADCs.
}
\end{figure}

The strategy of the experiment was to select a clean sample of muon stops,
$N_{\mu}$, in the fiducial volume of an active target well isolated from the
target walls and to ensure a 100\% efficiency in detection of the tritons,
$N_t$, from reaction (\ref{eq:reac1}). Then, the ratio
$N_t/N_{\mu}$ combined with the decay constant of the
muon will be a direct measure of the muon capture rate. 

The experiment was performed at Paul Scherrer Institute (PSI)
using the $\mu$E4 superconducting muon channel.
The apparatus is shown in Fig.\ref{FigSetup}. The target,
which is also the main detector, consists of a high pressure
ionization chamber (IC), filled with 120~bar $^{3}$He at room
temperature\footnote{The level of impurities was checked before
and after the run using a quadrupole mass spectrometer, and found
to be $8\times 10^{-5}$ for $^4$He and less than $3\times 10^{-5}$
for isotopes with $Z>2$. }. 
The ionization chamber includes a total of 14 anodes
(see Fig.\ref{FigAnode}a):
5 entrance anodes (type C),
5 stopping anodes (type B), one of which is split,
and 3 veto anodes (type A).
The anodes have a pad pattern, which allows the tracking and location of the
stopping muon. The B-anodes cover an area of $2.5\times 4.0\,{\rm cm^2}$.
The cathode to grid and grid to anode distances are $1.43\,{\rm cm}$ and
$0.08\,{\rm cm}$, respectively. This amounts to a sensitive volume of
$15\,{\rm cm^3}$. The cathode voltage is $-40\,{\rm kV}$, while the grid is
kept at $-3.5\,{\rm kV}$. These conditions result in a maximum electron drift
time of $3.2\,{\rm \mu s}$. The energy resolution is $30\,{\rm keV}$
($\sigma$). The IC is surrounded by an array of plastic scintillation counters
for the detection of neutrons from the breakup reactions and electrons from
muon decay. In the present experiment, this detector array serves
mainly for time scale calibration. A beam telescope of thin plastic
scintillators provides a fast trigger and also eliminates double-muon
events in a $\pm 8\,{\rm\mu s}$ time interval. The muons enter the IC through
a Beryllium window of thickness $4\,{\rm mm}$ with minimal Coulomb scattering
in the window.

A two-level charge-integrating trigger was used for the B-anodes. The three
main modes were $E_{\mu}$ (threshold $140\,{\rm keV}$, typical rate
$3000\,{\rm s^{-1}}$) which fires for each muon stopping over the B-anodes,
$E_{\mu-{\rm t}}$ (two time-separated thresholds at $140\,{\rm keV}$, typical
rate $10\,{\rm s^{-1}}$) which fires for each muon track followed by a separated
triton signal, and $E_{\rm t}$ (threshold $1.2\,{\rm MeV}$, typical rate
$11\,{\rm s^{-1}}$) which fires for most tritons and for
each triton piled up on the muon signal). $E_{\mu-{\rm t}}$ together with
$E_{\rm t}$ provide a 100\% effective trigger for the triton events. For
triggered events, the signals from all anodes were recorded over a period of
$10\,{\rm\mu s}$ with flash ADCs (Fig.\ref{FigAnode}b).
In total, about $4\cdot 10^8$
muons entered the fiducial volume triggering $E_{\mu}$. The $E_{\mu}$ trigger
was prescaled by various factors $k$ = 500, 1000, and 2000 to reduce the data
rate. Finally, $9\cdot 10^5$ prescaled $E_{\mu}$ and $1.2\cdot 10^6$
$E_{\mu-{\rm t}}$ (+$E_{\rm t}$) triggers were registered on tape during a
four weeks running period.

\section{Data Analysis}

There are two basic event types in this experiment, those in which the triton
was emitted late enough ($\Delta t\geq 500 {\rm ns}$) to have its signal cleanly
separated from that of the muon (see example in Fig.\ref{FigAnode}b) and
those in which the two signals overlapped.
The bulk of the  statistics (85\%) comes from the separated
events. However, to determine the muon capture rate, we should measure the
total number of tritons including pileups. There are two possibilities for
measuring the number of the pileups: direct counting of the pileup signals
or exponential extrapolation to $\Delta t = 0$ of the time distribution of the
triton signals separated in time, with $\Delta t=t_{\rm t}-t_\mu$, where
$t_{\rm t}$ and $t_{\mu}$ are the arrival times of the triton and the muon
signals on the IC anodes.

This data was analyzed using both of these techniques which
have different systematic effects and therefore cross-check each
other. We label these techniques as analysis A (inclusive of pileups)
and analysis B (separated events only)\cite{Bal95,Vor96}.
Furthermore, each of these analyses
was performed independently in a smaller fiducial volume and with
a substantially different approach to the systematics: analysis C 
(inclusive of pileups) and analysis D (separated events only)\cite{Pry96}.
Note that, for instance, analysis A and analysis B are
identical in all phases except in finding the pileup events.
In analyses B and D, $\Delta t$ was determined with a mid-pulse
timing technique, with some corrections for the geometrical effects of the
muon and triton angles.  
Analyses C/D used the region of muon stops from the center of B2 to the
center of B5, while analyses A/B used the whole region B1-B5.
Also, analyses C/D used less vertical space by more conservative cuts
of the muon drift times. Analysis A, by including pileup events,
using more B-anodes, and extending the drift volume gave the largest triton
statistics (1141263 events).

We should stress that special attention was paid in all of our analyses to
guarantee strictly identical detection efficiencies for muons with and without
tritons. Below, for brevity,  we present in detail only the results of
analysis A together with some remarks about and the final results of
analyses B/C/D.

\subsection{Selection of Muon Stops} 

The position of a muon stop along the beam direction was determined by the
energy deposits on the stopping anode B$_i$ and the preceeding one.
In analyses A/B the muon stop position was determined by finding the
most downstream anode of those anodes with the biggest number of signals
(generally two at most). In analyses C/D this position was determined by
following a contiguous muon track from a single C-anode hit.
Vertically, the fiducial volume was determined by cuts on the drift time
of muon tracks ($0.5\,{\rm\mu s}\leq t_{\mu}\leq 2.3\,{\mu s}$) which set
the capture location at least $2\,{\rm mm}$ from the
anode and grid, such that the triton track would not touch these electrodes.
Note that the triton range was $1.52\,{\rm mm}$ under the conditions of our
experiment. Laterally, the volume was set by rejecting events in which there
was a muon signal on one of the A-anodes within $\pm 50\,{\rm ns}$ from the muon
track. This restricted the penetration into the A-anodes to less than
$0.5\,{\rm mm}$. Furthermore, the number of muon
stops over the side anodes was kept low by triggering only on single,
central C-anodes, selecting mostly events in the middle of the chamber and
further reducing double muon events.

\begin{figure}[htb]
\begin{center} \mbox{\epsfysize=80mm\epsffile{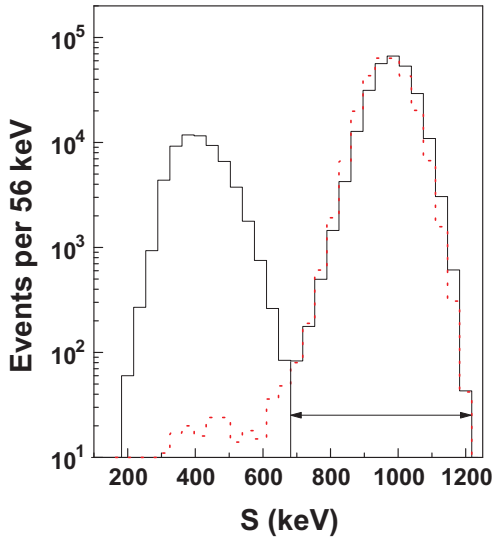}} \end{center}
\caption{\label{FigEDep}  
Distribution of muon energy depositions $S=E_i+2\cdot E_{i-1}$
with triggers $E_{\mu}$ (solid histogram) and $E_{\mu-{\rm t}}$ (dashed
histogram). The analysis window was set for the higher energy peak containing
the muons stopping on the anode, while the lower energy peak presents muons
that passed the anode.
}
\end{figure}

Finally, we strictly define the characteristics of the muon stop using
the muon energy deposits $E_i$ on the stopping anode $B_i$ and
$E_{i-1}$ on the previous anode. The distribution $S=E_i + 2\cdot
E_{i-1}$ (Fig.\ref{FigEDep}) shows a narrow peak for stopping muons.
By making a cut on $S$ we severely reject any track or noise except
a well defined muon stop. This cut is loose enough that we
estimate from the $S$-distribution less than 0.01\% difference in
efficiency for muons followed by capture tritons versus muons without
capture reaction.

\subsection{Detection of the Tritons}

In the analysis all triton events were subdivided into the following event
types:
\renewcommand{\labelenumii}{\arabic{enumi}.\arabic{enumii}}
\begin{enumerate}
\item 
Triton signal separated in time from the muon signal.
  \begin{enumerate}
  \item 
  Triton signal is on the same anode, $B_i$, as the muon stop.
  \item 
  Triton signal is split between $B_i$ and an anode neighbour to
  the one where the muon stopped.
  \item 
  Triton signal is completely contained on an anode adjacent to
  the one on which the muon stopped.
  \end{enumerate}
\item 
Triton signal overlaps with the muon signal. 
  \begin{enumerate}
  \item 
  Pileup signal is on the same anode, $B_i$, as the muon stop.
  \item 
  Pileup signal is split between $B_i$ and one of the
  neighbouring B- or A-anodes.
  \end{enumerate}
\end{enumerate}

\begin{figure}[htb]
\begin{center} \mbox{\epsfysize=80mm\epsffile{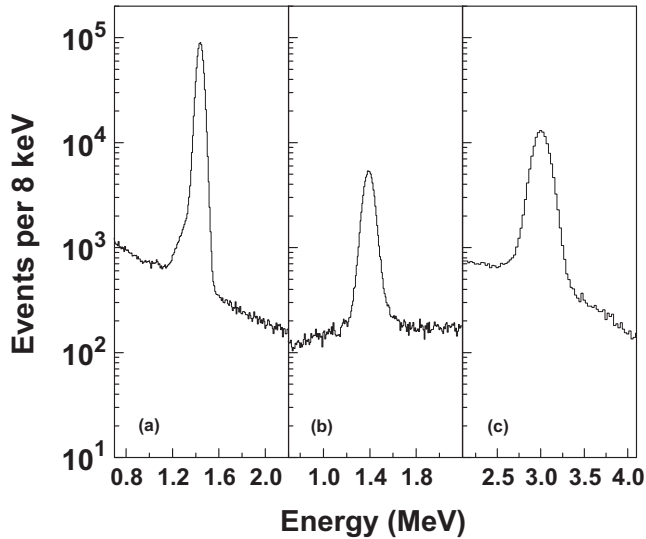}} \end{center}
\caption{\label{FigESpec}  
(a) Energy spectra of type~1.1 events (triton track contained
within a single B-anode), (b) of type~1.2 events (track not contained
within a single anode) and (c) of type~2.1 events (pileups contained
within a single anode). Note the vertical logarithmic scales.
}
\end{figure}

Fig.\ref{FigESpec} shows the energy spectra of the event types 1.1, 1.2 and 2.1,
which dominate the overall statistics. The
background under the triton peak is due to breakup reactions. 
In analysis A the background was subtracted by linear estimates based
on the background above and below the peaks. The systematic error was
then estimated by varying the regions of this linear fit.  Analyses C/D
used polynomial fits resulting in comparable errors. The
numerical results of analysis A are presented in Table \ref{tab1}.

\begin{table}[hbp]
\begin{center}
\begin{tabular}{|l|l|l|l|l|}
\hline
Event type  & $N_t$ & $N_t/N_t^{\rm tot}$ (\%) & BG &
BG/$N_t$ (\%) \\
\hline
1.1   & ~843322 & ~73.9 & 30882 & ~3.7 \\
\hline
1.2   & ~~~77881 & ~~6.8 & 11943 & 15.3 \\
\hline
1.3   & ~~~52638 & ~~4.6 & ~~2684 & ~5.1 \\
\hline
2.1   & ~157165 & ~13.8 & 20144 & 12.8 \\
\hline
2.2   & ~~~10257 & ~~0.9 & ~~9889 & 96.4 \\
\hline
sum & 1141263 & 100.0 & 75542 & 6.62 \\
\hline
\end{tabular}
\end{center}
\caption{Total statistics of analysis A. $N_t$ is the number of triton events
found on each anode, $N_t^{\rm tot}$ the sum, and BG is the number of evaluated
background counts underneath the triton line.}
\label{tab1}
\end{table}

Several corrections were applied to the measured number of $N_t$. The main
correction (6.445\%) is a trivial one taking into account the time beyond the
analyzed time interval of $5.995\,{\rm\mu s}$. Other corrections compensate 
mainly for rare event types, e.g. tritons passing into one of the A-anodes,
not listed in Table 1, and for small
losses of triton events. The sum of these corrections is 0.735\% (see
Ref.\cite{Vor96}), bringing the total correction to $N_t$ to
$\epsilon_t$ = 7.18\%.

\section{Results}

The calculation of the muon capture rate was done according to the formula
\begin{equation}
\lambda_{\rm stat}=\lambda_0\cdot\frac{N_t}{k\cdot N_{\mu}
(1-\epsilon_t)}
\label{eq:l_stat}
\end{equation}  
where $N_t$ is the number of detected tritons, $N_{\mu}$ is the number of
selected muon stops, {\it not} followed by muon capture, $\lambda_0 = 0.45516
\cdot 10^6\,{\rm s^{-1}}$ is the decay rate of the muon, and $k$
is the trigger prescaling factor. Note that in our experiment, the muon
stops which are not followed by muon capture are the events triggered by
$E_{\mu}$ only. This trigger contains practically none of the muon capture
events (except for about 10\% of low energetic breakup events giving a
negligible correction to $N_{\mu}$ of 1.4$\cdot 10^{-4}$).

The statistical error is determined by the following expression
\begin{equation}
\sigma_{\lambda_{\rm stat}} = \lambda_{\rm stat}\cdot
\sqrt{\frac{1}{N_t}+\frac{1-F}{N_{\mu}}}
\label{eq:sigma}
\end{equation}
where $F$ is the fraction of muon stops used in the analysis.
(In the approximation of Eq.~\ref{eq:sigma}
accounting for all muons would leave only
the statistical error of $N_t$.) In the
considered case of analysis A: $N_t$ = 1141263, $N_{\mu}$ = 349479,
$F$ = 0.356, while the total number of selected muon stops
corrected by the prescaling factor is $k\cdot N_{\mu}$ = 374028500.
Substituting these numbers for each $k$ value separately in (\ref{eq:l_stat})
and (\ref{eq:sigma}) and combining results statistically we obtain
\begin{equation}
\lambda_{\rm stat}=(1496.2\pm 2.6_{\rm stat}) {\rm s}^{-1}.
\end{equation}

The systematic error was evaluated by varying the energy and
time cuts in wide ranges and comparing results from independent data on
different anodes and under different running conditions.
The main contribution to the total systematic error comes from
uncertainties in the background subtraction (0.16\%).  The residual
systematic errors in the determination of $N_\mu$ and $N_t$ are
0.03\% and 0.05\% respectively, giving a total systematic error of
0.17\%.

\begin{table}[hbp]
\begin{center}
\begin{tabular}{|l|l|l|l|l|}
\hline
 $k$  & ~~~2000 & ~~~2000 & ~~~1000 & ~~~~500 \\
\hline
$N_{\mu}$   & ~~53664 & ~~70619 & ~~25729 & ~199467 \\
\hline
$N_t$   & ~327252 & ~429989 & ~~78622 & ~305400 \\
\hline
B1   & 1497$\pm$16 & 1500$\pm$15 & 1514$\pm$23 & 1509$\pm$11 \\
B2   & 1483$\pm$15 & 1489$\pm$14 & 1494$\pm$21 & 1482$\pm$10 \\
B3   & 1484$\pm$15 & 1471$\pm$13 & 1482$\pm$20 & 1503$\pm$10 \\
B4   & 1498$\pm$14 & 1486$\pm$13 & 1514$\pm$20 & 1516$\pm$10 \\
B5   & 1506$\pm$14 & 1525$\pm$13 & 1497$\pm$20 & 1492$\pm$10 \\
\hline
B1-B5   & ~1493$\pm$6\hphantom{0} & ~1495$\pm$5\hphantom{0} &
 ~1501$\pm$8\hphantom{0} & ~1500$\pm$4\hphantom{0} \\
\hline
\end{tabular}
\end{center}
\caption{$k, N_\mu, N_t$ and partial results on $\lambda_{\rm stat}$ from
different B-anodes and run groups}
\label{tab2}
\end{table}

Table \ref{tab2} presents the partial results in measuring $\lambda_{\rm stat}$
(analysis A). These results show a remarkable stability.
The final result from analysis A is
\begin{equation}
{\rm analysis~A:~~~~~~~~~~} \lambda_{\rm stat} = (1496.2 \pm 2.6_{\rm stat}
\pm 2.6_{\rm syst}) {\rm s^{-1}} \\
\end{equation}
The results from the other three analyses are:\\
\begin{equation}
{\rm analysis~B:~~~~~~~~~~} \lambda_{\rm stat} = (1495.8 \pm 3.0_{\rm stat}
\pm 2.8_{\rm syst}) {\rm s^{-1}} \\
\end{equation}
\begin{equation}
{\rm analysis~C:~~~~~~~~~~} \lambda_{\rm stat} = (1497.4 \pm 3.1_{\rm stat}
\pm 3.8_{\rm syst}) {\rm s^{-1}} \\
\end{equation}
\begin{equation}
{\rm analysis~D:~~~~~~~~~~} \lambda_{\rm stat} = (1496.0 \pm 4.5_{\rm stat}
\pm 3.0_{\rm syst}) {\rm s^{-1}} \\
\end{equation}

Let us recall, that the principal difference between analyses B/D and A/C
is that B/D used only the separated events, requiring extrapolation back to
$\Delta t = 0$. 
This method relies on the precision of the
determination of the absolute value of the $\Delta t = 0$ position.
We have shown that the mid-pulse to mid-pulse technique determines
this position with a precision better than $\pm$2 ns.  In analysis B
this was proven using the neutron detectors to compare the arrival times
of decay electrons with neutrons from the breakup channels and correcting
for the estimated time of flight.  In analysis D, $\Delta t = 0$ was confirmed
by checking the average angle and curvature of muon tracks to estimate
any offset to the mid-pulse technique. In all analyses, several other
systematic effects were checked.  These include: edge effects, binning
errors, triton track angle effects, and pedestal shift effects.
The very good agreement among all analyses demonstrates the low level
of remaining systematic errors.

Finally, we present our result as\\
\begin{equation}
\lambda_{\rm stat} = (1496.0 \pm 4.0) {\rm s^{-1}} \\
\end{equation}
where the error is the combined statistical and systematic error of
the larger statistics analyses A and B.

\begin{figure}[htb]
\begin{center} \mbox{\epsfysize=80mm\epsffile{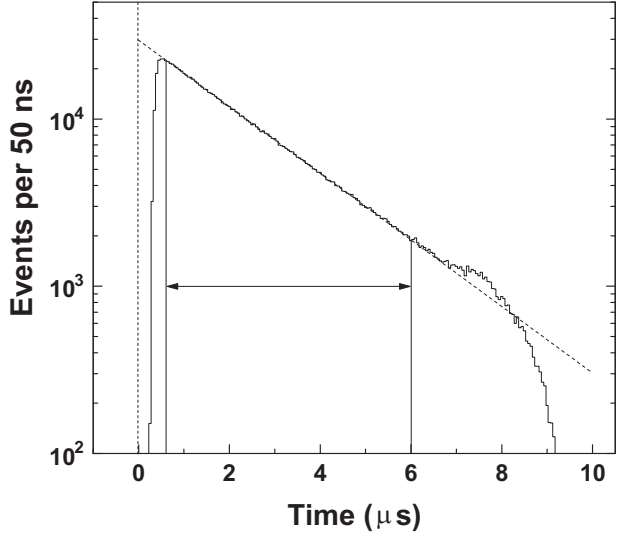}} \end{center}
\caption{\label{FigT}     
Time distribution of separated $\mu-{\rm t}$ events. The events
near $t=0$ are absent due to muon-triton signal overlap. In the time region
for analysis B $(0.6-6.05\,{\rm\mu s})$ a pure exponential slope with
$\lambda=(0.4573\pm0.0005)\,{\rm s^{-1}}$ was found, in accordance with the
known lifetime of the $\mu^{3}$He muonic atom.
}
\end{figure}

The measured value of $\lambda_{\rm stat}$ corresponds to a statistical
population of the singlet and triplet states of $\mu^3$He atoms. An assumed
hyperfine transition from the singlet to the energetically lower triplet state
with a rate $\lambda^{0\to 1} = 0.01\,{\mu s^{-1}}$ would change the
observed $\lambda_{\rm stat}$ by +0.17\%. Fortunately, we can limit this
process in our experiment by comparing the observed time distribution
(Fig.~6) of the tritons with the expected one:
\begin{eqnarray}
  \frac{dN_t}{dt} & = & 
  \frac{1}{4}\lambda^0_{\rm H} e^{-(\lambda_0+\lambda^0_c+\lambda^{0\to 1})t}
 +\frac{3}{4}\lambda^1_{\rm H} e^{-(\lambda_0+\lambda^1_c)t} + \nonumber
 \\ &   & 
 +\frac{1}{4}\frac{\lambda^{0\to 1}\cdot\lambda^1_{\rm H}}
                  {\lambda^{0\to 1}+\lambda^0_c-\lambda^1_c}
  (e^{-(\lambda_0+\lambda^1_c)t} -
   e^{-(\lambda_0+\lambda^0_c+\lambda^{0\to 1})t}) \ \ .
\end{eqnarray}
Using the rate
$\lambda_{\rm stat}=(\lambda^0_{\rm H}+3\lambda^1_{\rm H})/4$
determined above and the total spin averaged capture rate 
$(\lambda^0_c+3\lambda^1_c)/4=(2216\pm 70)\;{\rm s}^{-1}$ \cite{Maev96}
together with the theoretical values for the differences between the
singlet and triplet capture rates
$\lambda^0_{\rm H}-\lambda^1_{\rm H} = 578 \;{\rm s}^{-1}$\cite{Con92} and 
$\lambda^0_c-\lambda^1_c = 450 \;{\rm s}^{-1}$\cite{Con94}
we obtain
\begin{equation}
\lambda^{0\to 1} = (0.006 \pm 0.008)~ \mu{\rm s}^{-1} \\
\end{equation}
thus restricting the possible influence of this process to
a level of $\leq$0.14\%.

Another uncertainty might arise if a significant fraction of the 2s-state
were metastable with a lifetime comparable to the muon decay rate. The
total initial population for the 2s-state is 7\%\cite{Rei87}.
According to the present understanding\cite{Rei87,Eck86}, however,
the quenching of the 2s-state occurs on a much shorter time scale and does not
influence our result. But again, we can limit
this effect with our own data by comparing the number of tritons, $N_t$,
obtained by the extrapolation method (analyses B/D) with that obtained
by direct counting of pileups (analyses A/C). From this comparison we obtain
the upper limit for the lifetime of the 2s-state
\begin{equation}
\tau_{\rm 2s} \leq 50~ {\rm ns}.
\end{equation}
The effect of this transition is indeed negligible.

\section{Discussion}

The obtained result can be analyzed in the framework of the EPM\cite{Prim65}.
In this model, the weak current of the $^3$He-$^3$H transition is parametrized
by four form factors $F_{\rm V}$, $F_{\rm M}$, $F_{\rm A}$, and $F_{\rm P}$,
evaluated at the relevant value of the four-momentum transfer
$q_0^2=-0.954m_\mu^2$, see e.g.
Ref.\cite{Con92}. Two of these parameters, the vector and magnetic form 
factors, $F_{\rm V}$ and $F_{\rm M}$, are
derived from the conserved vector current (CVC) theorem and from
the results of electron elastic scattering by $^3$He and $^3$H\cite{Con92}:
\begin{equation}
  F_{\rm V}(q_0^2)=0.834\pm0.011, \ \ \ \
  F_{\rm M}(q_0^2)=-13.969\pm0.052.
\nonumber
\end{equation}
Taking these data as inputs, our measurement determines the correlation
between the pseudoscalar form factor $F_{\rm P}(q_0^2)$ and the axial form
factor $F_{\rm A}(q_0^2)$, as shown in Fig.~7.
The width of the allowed region
is determined mostly by the error in $F_{\rm V}(q_0^2)$.
In a further step we use $F_{\rm A}(q_0^2) = 1.052 \pm 0.010$
extrapolated\cite{Con92} from $F_{\rm A}(0) = 1.212 \pm 0.005$ as determined
from the $^3$H beta decay rate. Note that this extrapolation involves some
theoretical consideration based on the impulse approximation that results in
the increase of the error in $F_{\rm A}(q_0^2)$. This procedure allows
determination of the pseudoscalar form factor
\begin{equation}
F_{\rm P}(q_0^2)=20.8\pm 2.8.
\end{equation}
The main contribution to the error comes from the error on $F_{\rm A}(q_0^2)$,
while the contribution from our measurement is only $\pm$~0.5~.

\begin{figure}[htb]
\begin{center} \mbox{\epsfysize=80mm \epsffile{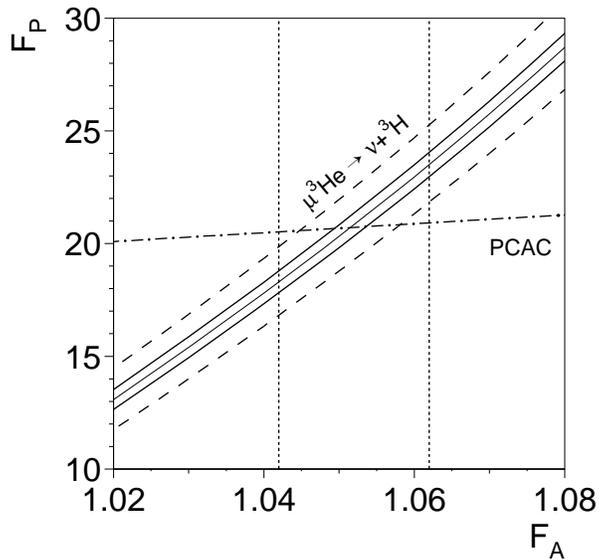}} \end{center}
\caption{\label{FigFAFP} 
Plot of axial form factor $F_{\rm A}$ versus pseudoscalar form
factor $F_{\rm P}$ showing the constraints by the experimental errors of our
measurement (full lines) and with errors in $F_{\rm V}$ and $F_{\rm M}$ added
(dashed lines). The vertical dotted lines give the constraints on $F_{\rm A}$,
and the dot-dashed line presents the PCAC prediction.
}
\end{figure}

The obtained value of $F_{\rm P}$ can be compared with the PCAC
prediction that relates the pseudoscalar and axial form factors:
\begin{equation}
F_{\rm P}^{\rm PCAC}(q^2)=\frac{m_\mu(M_{^3{\rm He}}+M_{^3{\rm H}})}
{m^2_\pi-q^2}\,F_{\rm A}(q^2)\ \,
\label{eq:FPPCAC}
\end{equation}
\noindent
This relation is shown in Fig.\ref{FigFAFP}
by the dot-dashed line. Taking again the
extrapolated value for $F_{\rm A}$, one obtains
\begin{equation}
F_{\rm P}^{\rm PCAC}(q_0^2)=20.7\pm 0.2
\end{equation}
\noindent
in perfect agreement with our experimental result (15).

The availability of precise three-body wave functions of the $A=3$ system
makes it possible to perform calculations of the
muon capture rate on a ``microscopic'' level. Calculations done in the impulse
approximation lead to a capture rate\cite{Con92} of $\lambda_{\rm stat}=
1304\,{\rm s^{-1}}$ which deviates from the observed rate by 13\%.
This clearly indicates that MEC must be accounted for.
Full microscopic calculations including MEC have been
performed recently by Congleton and Truhl\'{\i}k\cite{Con96}. The authors
obtained  $\lambda_{\rm stat}=1502\pm 32\,{\rm s}^{-1}$ in excellent agreement
with our experimental result (10). This agreement can be interpreted
as a confirmation of the validity of the approach used to include the MEC in
the microscopic theory.

Note that these calculations required the knowledge of the nucleon form
factors $g_{\rm V}$, $g_{\rm A}$, $g_{\rm M}$, and $g_{\rm P}$ at $q_0^2$.
Three of these, namely $g_{\rm V}$, $g_{\rm M}$, and $g_{\rm A}$, are
experimental values, while the pseudoscalar form factor $g_{\rm P}$
was determined by the PCAC relation\cite{Adl66} to be
\begin{equation}
g_{\rm P}^{\rm PCAC}(q_0^2)=8.12~.
\end{equation}
In this context, we wish to point out two recent publications\cite{Mei94,Fea97}
where, for the nucleon, a formula identical to the PCAC relation was derived
within the framework of chiral perturbation theory.
Assuming the validity of the microscopic model within the errors cited in
Ref.\cite{Con96}, we can determine $g_{\rm P}(q_0^2)$
from comparison of the calculated value of $\lambda_{\rm stat}$ with our
experimental result:
\begin{equation}
g_{\rm P}(q_0^2)=8.53\pm1.54.
\end{equation}
\noindent
The error of this result is largely dominated by the errors in
the calculation of the theoretical muon capture rate and {\it not} in the
precision of the present experimental result. Note also that our value of
$g_{\rm P}(q_0^2)$ agrees well with the PCAC prediction
\begin{equation}
\frac{g_{\rm P}(q_0^2)}{g_{\rm P}^{\rm PCAC}(q_0^2)}=1.05\pm0.19.
\end{equation}
while the recent experiment\cite{Jon96} on radiative muon capture on hydrogen
shows a significant deviation. Further studies of muon capture on hydrogen are
definitely needed for a determination of $g_{\rm P}$ with higher accuracy.\\

\section*{Acknowledgement}

This work was supported in part by the Russian Ministry of Sciences, the
Austrian Science Foundation, the German Federal Ministry of Research and
Technology and the Paul Scherrer Institute.
The accelerator and support staff of the Paul Scherrer Institute
is gratefully acknowledged for their efficient help and expertise,
without which the experiment could not have been a success.


\end{document}